\documentstyle[12pt,epsf]{article}
\input epsf
\voffset = - 0.3in
\hoffset = - 0.5in
\def\lagr{\hbox{$\cal L$}}

\def\dalam{\hbox
{\vrule\vbox{\hrule\hbox to 1ex{ \hfill}\kern 1 ex\hrule}\vrule}}

\def\half{\hbox{$ {1 \over 2}$ }}
\def\i/h{{i \over \h}}
\def\1/2{\hbox{$ {1 \over 2}$ }}

\def\vf{\varphi}
\def\f{\phi} 
\def\p{\psi}
\def\bp{\bar \psi}
 \def\E{\hbox{$\cal E $}}
\def\ve{\varepsilon}

\def\<{\langle}
\def\>{\rangle}

\def\ch{\cosh}
\def\sh{\sinh}

\def\h{\hbar}

\def\a{\alpha}
\def\b{\beta}
\def\g{\gamma}  
\def\d{\delta}  
\def\l{\lambda}   
\def\s{\sigma}
\def\r{\rho}  

\def\c{\chi}

\def\m{\mu}
\def\n{\nu}

\def\w{\omega}
\def\tt{\theta}

\def\({\left(}
\def\[{\left[}
\def\){\right)}
\def\]{\right]}

\def\pd{\partial}

\title{\bf Chiral Hybrid Bag Model with the Boson Field
Inside the Bag} \author{I.
Cherednikov$^{a,b}$ \thanks{e-mail address:  igorch@goa.bog.msu.su }
, \ S. Fedorov$^{a}$, \ M. Khalili$^{a}$, \\ and K.
Sveshnikov$^{a,c}$  \\ {\em $^{a}$ Physics Department, Moscow State
University, Moscow, Russia} \\ {\em $^{b}$  The Abdus Salam 
ICTP, Trieste, Italy} \\ {\em $^{c}$
 Institute for Theoretical Problems of Microphysics} \\ 
{\em Moscow State University, Moscow, Russia} }

\begin{document}

\maketitle

\rightline{Preprint ICTP IC/99/191}

\begin{abstract}
\noindent The three-phase version of the hybrid chiral bag model,
containing the phase of asymptotic freedom, the hadronization phase
as well as the intermediate phase of constituent quarks, is proposed.
For this model the self-consistent solution,
which takes into account  the fermion vacuum
polarization effects, is found in (1+1) D.  Within this solution
 the total energy of the bag, including the one-loop
contribution from the Dirac's sea, is studied as a function of
bag geometry under condition of  nonvanishing boson condensate
density in the interior region. The existence and
uniqueness of the ground state bag configuration, which minimizes
the total energy and contains all the three phases, are shown.

\end{abstract}

\vskip 1.0in


\noindent

\vfill

\eject

\section{Introduction}

The idea to describe a hadron as a bounded region of space
(``the bag '') filled with quark and gluon fields appears nowadays
to be one of the most natural ways of constructing the effective
quantum field theory with total confinement of  color objects
[1-12].  The boundary conditions are chosen as to
confine the colored particles to the bag without breaking the
relativistic covariance [1-3]. The very possibility of existence of
such an object is related to the change of the vacuum structure
inside the hadron. It is supposed that the non-perturbative vacuum
inside strongly interacting particle is completely destroyed, what
produces an excess of energy proportional to the bag volume [1-5].
The stability of this system is provided by the valence quarks
contribution to the total energy proportional to the inverse bag
radius.  Even the early MIT-bag model yielded consistent results for
the mass spectrum and other static characteristics of hadrons with
relatively small set of free parameters [6-8]. Further development of
this model has led to taking into account the effects of spontaneous
chiral symmetry breaking --- one of the most important features of
strong interactions at low energies, and to incorporating the meson
fields into the theory which play the role of Goldstone bosons
[9-11].  (Depending on the number of light-quark flavors, it is
either the triplet of $\pi$-mesons for SU(2) group or the octet of
$\pi$- and $K$-mesons for SU(3)).  As a result the most consistent
approach to the description of hadronic structure based on bag
models has been elaborated within the framework of so-called Hybrid
Chiral Models (HCM) [10-12].  The HCM treat the nucleon as the
little bag with quarks and gluons confined inside, surrounded by a
big cloud of virtual mesons. The latter can be described either by
the ``hedgehog'' $\pi$-meson solution of the Skyrme model, or by some
more complicated models including vector mesons [13-15].

On the whole, the models of such a type predict correct scales for
various static characteristics of non-strange baryons. At the same
time, the estimation of certain values may yield essential
quantitative discrepancies, even the sign may be incorrect [12,16].
Moreover, the best HCM results have been obtained for such values of
model free parameters (which are the vacuum pressure constant $B$,
the coupling constant $\a_s$, the current quark masses, and the
constant of Casimir energy $Z$), that differ from their values
obtained by means of other methods [4,5]. These
circumstances indicate that the current formulation of HCM suffers
from a number of shortcomings.

Actually, the most subtle point in the HCM is the Cheshire Cat
Principle (CCP) which is basic for parting space into regions with
different phases inside of them [17]. The essence of CCP is the
hypothesis [18] that the fermion theory inside of bag and the boson
theory outside are actually equivalent and can be transformed into
each other via the bosonization procedure. Therefore  none of
physical properties of such bag  depends actually on the choice of
boundary surface, while the boundary conditions are determined from
the bosonization equations [19].  However, the bosonization, as well
as CCP, can be rigorously proved only in (1+1)D, while in the real
(3+1)D world the solution for the bosonization problem is still
absent.  As a result, in the (3+1)-dimensional HCM based on CCP there
exists a rather small set of observables (e. g.,  the topological
charge), which do not really depend on the bag radius [20]. Moreover,
the phenomenology of strong interactions  predicts unambiguously the
existence of the characteristic confinement scale about 0.5 fm, and
so in the realistic (3+1)D models the CCP should be strongly
violated, regardless on the proof of bosonization.

Therefore it seems to be natural to modify HCM as to avoid any relation to
CCP and specific features associated with, such as an infinitely
thin boundary surface between different phases and corresponding
surface action.  Such a modification appears to be possible provided
the correlation between the different phases of bag, which are
now not assumed to be equivalent in the sense of bosonization, arises
due to some interaction of the real status taking place in a finite
region of space.  The latter can be naturally treated as the region
of the intermediate third phase of the bag [21].  The introduction of
such an additional phase is useful also because it allows to
introduce the chirally invariant mechanism of the quark mass
generation, hence it can be considered to a certain extent as the
phase of constituent quarks, while the original 2-phase model does
not take into account their existence at all.  Meanwhile, the concept
of constituent quarks has been shown to  be very efficient in the
hadron spectroscopy.  From this point of view  the most attractive
physically situation is the one, where the initially free,
almost massless quarks (corresponding to large momentum transfer)
transmute firstly into ``dressed'' due to interaction massive
constituent quarks carrying the same quantum numbers of color,
flavor, and spin, and only afterwards there emerges the purely
mesonic colorless phase.

The first order approximation to such version of the bag is given by
the 3-phase hybrid model with additional constituent quark phase instead of
the boundary with vanishing radial extent [21]. This model allows
to take into account three phases: the phase of asymptotic freedom
with free massless quarks, the phase of constituent quarks which
acquire an effective mass due to the chirally invariant interaction
with the boson field in the intermediate region of finite size, and
the hadronization phase where the creation of free quarks is
suppressed by infinitely large mass, while the non-linear dynamics of
boson field causes the appearance of the c-number boson condensate in
the form of a classical soliton solution, which ultimately accounts
for the quantum numbers of the whole bag.

In the present paper the toy model of such kind in (1+1)D is
considered, where a single-flavor fermion field is coupled in a
chirally invariant way to the real scalar field possessing a
non-linear soliton solution in the exterior region. The
self-consistent solution to the model equations, which takes into
account the effects of fermion-vacuum polarization, is found.  Within
this solution the renormalized total energy of the
bag is studied as a function of the bag geometry under condition
of nonvanishing boson field in the interior region.
It is shown that by suitable choice of the model
parameters the configuration with
minimal total bag energy and containing all three phases, exists
and is unique.

\section{Lagrangian and Equations of Motion}

The division of space into phases is performed by means of the
system of subsidiary fields $\tt(x)$ according to the method of ref.
[3, 22].  The essence of this method may be explained  as
following.  Let us consider the Lagrangian of the form:
$$\lagr_0 = \half (\pd_{\mu}\f)^2 - \tt V(\f) + \half
(\pd_{\mu}\tt)^2 - g_{0}^2 W(\tt), \eqno(2.1)$$ where the
 coupling constant $g_{0}$ of self-interaction of field $\tt$ is
 assumed to be large enough to neglect the matter fields
 $\f$ in the dynamics of $\tt$ to the leading order, and thereafter
 to use  $\tt$  as background fields for
the dynamics of $\f$'s [21,22].  One can obviously construct the
Lagrangian containing as many fields $\tt (x)$ as needed with
appropriate self-interaction, which will determine (almost)
rectangular division of space into regions corresponding to different
phases, while the Lorentz-covariance will be broken only
spontaneously, namely on the level of solutions of equations of
motion.  The latter circumstance allows one to use the framework of
covariant group variables [23] in order to restore the covariance.
Assuming further that subsidiary fields $\tt (x)$ have already
formed the required bag configuration, let us start with the
following Lagrangian:  $$\lagr = \bp i\hat\pd \p +
\half(\pd_{\mu}\vf)^2 -\tt(x_1 < |x| < x_2) \left({M \over 2}
\left[\bp, e^{ig \g_5 \vf} \p \right]_{-}\right) - $$ $$
-\tt(|x|>x_2) \left({M_0 \over 2} \left[\bp, e^{ig \g_5 \vf}\p
\right]_{-}+V(\vf)\right).  \eqno(2.2)$$ The commutator of fermion
fields in terms, describing the chiral fermion-boson coupling,
ensures the charge conjugation symmetry of the model.

So in initial stage we have  the theory of two fields,
the spinor field $\p$ and the boson field $\vf$. These fields are free and
massless in the region $I: \{ |x|<x_1 \} $;
in the region $II:  \{x_1
\leq |x| \leq x_2 \} $ the field $\vf$ interacts in a chirally
invariant way with fermions resulting in emergence of the effective
fermion mass $M$; and in the region  $III:  \{ |x|>x_2 \} $ the effective
fermion mass increases up to $M_0$ and the self-interaction
of $\vf$ switches on, providing the appearance of a soliton
solution for the boson field in this region.  Note that in this model
the vacuum pressure term appears to be redundant, since due to the
existence of the intermediate phase the Dirac's sea polarization
behaves very specifically and ensures itself the required ``inward
pressure''.  Moreover, in our model there is no special need in
``valence'' fermions, since it is the boson condensate in the form of
the topological soliton, which accounts now for  all the bag's
quantum numbers, represented by  the topological ``baryon number'' in
this simplest case.  The latter point makes it difficult to explain
the need for the vacuum pressure term in this model too.

In order to provide the confinement of fermions we assume the mass
$M_0$ to be very large, what leads to the dynamical suppressing
the fermion field in the exterior region $III$. At the same time, in
the region of asymptotic freedom we have free (decoupled from
fermions) massless scalar field and so the nonvanishing
density of the boson condensate.  This situation does not contradict
the general concept of bag models, and can be treated as one of
possible versions of the 3-phase model. Another possibility with
vanishing scalar field in inner region has been studied in
detail in ref.  [21].  Let us assume also, that the solution of
equations of motion for the boson field is an odd topological
soliton.   The case of an even soliton with nonvanishing scalar field
inside is of no interest, since in this case the trivial solution
(scalar field is a constant equal to one of minima of
self-interaction potential $V(\vf)$) is the only energetically
preferable.

Let us consider now the behavior of fields in detail.
According to the general approach accepted in hybrid models we
consider the boson field in the mean-field approximation, i. e. it is
assumed to be a $c$-number field.  Neglecting temporarily the
explicitly Lorentz-covariant description, we'll consider the
center-of-mass system of the bag, where $\vf(x)$ is stationary
classical field being a background for evolution of fermions.
The equations of motion read:

\noindent in the region $I$
$$ i\hat \pd \p=0, \eqno(2.3a)$$
$$ \vf ''  = 0, \eqno(2.3b)$$

\noindent in the region $II$
$$ \left(i\hat\pd - M e^{ig \g_5\vf } \right)\p = 0, \eqno(2.4a)$$
$$ \vf ''= ig {M \over 2} \langle\left[\bp, \g_5 e^
          {ig\g_5\vf}\p\right]_{-}\rangle, \eqno(2.4b)$$

\noindent and in the region $III$
$$ (i\hat \pd  - M_0 e^{ig \g_5\vf } )\p = 0, \eqno(2.5a)$$
$$ -\vf''+ V'(\vf) = 0 \ , \eqno(2.5b)$$
where $\< \ \ \ \>$ in eq. (2.4b)
stands for the expectation value in a given bag state.
To simplify calculations, we put further $g=1$,
because the dependence on it can be easily restored by means of
the substitution $\vf \to \vf/g$.
Then the spectral problem for fermionic wave functions
$\p_{\w}$ with definite energy $\w$ reads  $$ \w \p_{\w} = -i\a \p
'_{\w}+ \b e^{i\g_5\vf}\left[M \tt (x_1 < |x| < x_2)+M_0\tt(|x|>x_2)
\right]\p_{\w}.  \eqno(2.6)$$
To complete the formulation of the spectral problem, the
following boundary conditions have to be imposed $$\pm
i\g^1 \p_{\w}(\pm x_2) + e^{i \g_5 \vf(\pm x_2) } \p_{\w}(\pm x_2) =
0 \ , \eqno(2.7)$$ provided with the condition of continuity
for $\p (x)$ on boundaries separating the regions $I$ and $II$.
Note that the boundary conditions (2.7) are actually the
standard chiral boundary conditions for the hybrid models [9-12].
However, they arise now as a direct consequence of an infinite mass of
fermions in the region $III$, rather than from the local surface
action, what is not completely correct [21].  In the region $I$ the
equation (2.6) is the equation for free massless fermions $$ \w
\p_I=-i\a \p'_I \ , \eqno(2.8)$$ while in the intermediate region
$II$ one has $$ \w \p_{II}=-i \a \p'_{II} + \b M e^{i\g_5 \vf}
\p_{II} \ .  \eqno(2.9)$$ Conditions of wavefunction's continuity on
the boundary between I and II are given by $$ \p_{I}(\pm x_1)=
\p_{II}(\pm x_1) \ , \eqno(2.10)$$ while at points $|x|=x_2$ the
wavefunctions satisfy the conditions (2.7).  By this the field $\vf$
in eq. (2.9) is not arbitrary but has to be determined
self-consistently from eq.  (2.4b) with appropriate
continuity conditions imposed on the field and it's derivatives at
points $|x|=x_{1,2}$.

\section{Self-consistent solution of bag equations}

The essential feature of this bag configuration is the fact, that
the self-consistent equations (2.4) in the closed intermediate region
$II$ of finite size $d=x_2-x_1$ possess simple and physically
meaningful solution, which would be unacceptable if these equations
were considered in the infinite space. In order to obtain this
solution, we perform in the region $II$ the chiral Skyrme rotation $$
\p = \hbox{exp}(-i\g_5\vf /2) \c, \eqno(3.1) $$ hence the equation
(2.9) and the boundary condition (2.7) transform into $$
(\w-\half\vf')\c_{\w} = -i\a \c'_{\w}+\b M\c_{\w}, \eqno(3.2)$$ $$
\pm i \g^1\c_{\w}(\pm x_2) + \c_{\w} (\pm x_2)=0 \eqno(3.3) $$
correspondingly.

It follows from (3.2) and (3.3) that if we assume the linear behavior
of the scalar field in the region $II$, namely  $$\vf' = \hbox{const}
= 2\l, \eqno(3.4)$$ then the equation (3.2)  becomes the equation for
free massive fermions:  $$ \n \c =-i \a \c' + \b M \c  \eqno(3.5)$$
with eigenvalues $\n=\w-\l$.  Therefore, as we expected, the massless
(in the region $I$) fermions acquire the mass $M$ in the region $II$
due to the coupling to field $\vf$, and so the intermediate phase
emerges which describes massive quasifree ``constituent quarks''.
Continuity conditions for the scalar field, which should be an odd
function, lead us to the unique solution $$ \vf(x)=2\l x \ .
 \eqno(3.6)$$ The equation (3.5) obviously possesses the sign
symmetry $\n \to -\n$, which corresponds to the unitary
transformation of fermionic wave function $$ \c \to \tilde \c = i
\g_1 \c \ .  \eqno(3.7)$$ The important point here is, that the axial
currents $$ j_5=i \bp \g_5 e^{i\g_5 \vf} \p =i \c^+ \g_1 \c
\eqno(3.8)$$ coincide for these sign-symmetric states $$ j_5=i \c^+
\g_1 \c = i {\tilde \c}^+ \g_1 \tilde \c = \tilde j_5 \ .
\eqno(3.9)$$ In general, however, one cannot derive the sign symmetry
for the fermion spectrum in our case from the sign symmetry $ \n
\leftrightarrow -\n $ in eq. (3.5), because the latter is true only
in the region $II$, while the spectrum has to be determined from the
Dirac equation on the unification of the regions $I+II$. The
straightforward solution of equations (2.8-9) with account of
boundary conditions (2.7) and constraint (2.10) gives the following
equation for spectrum:  $$\exp(4i\w x_1-2i\vf_1)=\frac{1-e^{-2i
kd}\frac{M-i(\n + k)}{M-i(\n - k)}} {1-e^{-2ik d}\frac{M+i(\n -
k)}{M+i(\n + k)}} \frac{1-e^{2ik d}\frac{M-i(\n - k)}{M-i(\n + k)}}
{1-e^{2ik d}\frac{M+i(\n + k)}{M+i(\n -k)}}, \eqno(3.10)$$ where
$\n^2=k^2+M^2 $ and $ \vf_1=\vf(x_1)$.  Analysing the equation (3.10)
one easily finds that the fermionic spectrum reveals the symmetry $
\n \leftrightarrow -\n $, if $$ 4\l x_1 - 2 \vf_1 =\pi s \ ,
\eqno(3.11)$$ where $s$ is integer, since for such values of the
derivative of the field $\vf (x)$ the l.  h.  s.  of equation
(3.10) reduces to $(-1)^s \ \exp (4i\n x_1) $. However, in our case
$\vf_1 =2\l x_1$,  hence the equation (3.11) leads to the
single possibility $s=0$, while the parameter $\l$ remains arbitrary.
This result differs crucially from the case considered in [21], where
$\vf_1= 0$ by virtue of vanishing of the boson condensate in the
region $I$, and so the set of solutions with different $s\not= 0$
emerges. In the latter case, however,  the equation (3.11) produces
the non-trivial relation between $\l$ and $x_1$.

According to eq. (2.4b), in the region $II$ $\vf''(x)$ is determined by
the v.e.v. of the  $C$-odd axial current
$$ J_5=\1/2 \[ \bp , i\g_5 e^{i\g_5 \f} \p \]_-=\1/2 \[ \c^+ , i\g_1
\c \]_- \ , \eqno(3.12)$$ with $\c$ being now the
secondary-quantized Dirac field in chiral representation (3.1) $$ \c
(x,t)= \sum \limits_n  b_n \c_n(x) e^{-i\w_n t} \ , \eqno(3.13)$$
where $\c_n(x)$ are the normalized solutions of the corresponding
Dirac equation, and $b_n \ , b^+_n$ are fermionic
creation-annihilation operators, which obey the canonical commutation
relations $$ \{ b_n , b^+_{n'} \}_+= \d_{nn'} \ , \quad  \{ b_n ,
b_{n'} \}_+=0 \ .  \eqno(3.14) $$ The average over the given bag's
state includes, by definition, the average over the filled sea of
negative energy states $\w_n <0 \ + $ possible filled valence fermion
states with $\w_n >0 $ , which are dropped for the moment because
their status is discussed specially below.  Finally,  $$ \<
J_5 \>=\< J_5 \>_{sea}= \( \1/2 \sum \limits_{\w_n <0} - \1/2 \sum
\limits_{\w_n >0} \) \ \c^+_n i\g_1 \c_n  \ .  \eqno(3.15)$$
Let us emphasize, that in (3.15) the division of fermions into sea
and valence ones is made in correspondence with the sign of their
eigen-frequencies $\w_n$, which differ from sign-symmetric $\n_n$
by the shift in $\l$ $$ \w_n=\n_n+ \l \ , \eqno(3.16)$$ and so do
not possess the symmetry $ \w \leftrightarrow -\w $.  However, if we
suppose additionally that $\n_n$ and $\l$ are such that for all $n$
the signs of $\n_n$ and $\w_n$ coincide, i. e. after shifting by
$\l$ none of $\n_n$'s changes its sign, then the condition $ \w_n {>
\atop <}  0$ in eq. (3.15) will be equivalent to the condition $ \n_n
{> \atop <}  0$ . Hence $$ \< J_5 \>_{sea}= \( \1/2 \sum
 \limits_{\n_n <0} - \1/2 \sum \limits_{\n_n >0} \) \ \c^+_n i\g_1
\c_n =0  \eqno(3.17)$$ by virtue of relation (3.9).  In turn,  it
means that the eq. (2.4b) in the region $II$ reduces to
$\vf''=0$, what is in excellent agreement with the
assumption that $\vf'(x)=const$ in the region $II$.  In other words,
we obtain the solution of coupled equations (2.4) in the region $II$
in the form of the linear function (3.6) for the scalar field, and
the equation (3.16) for the fermion energy spectrum, where $\n_n$ is
defined from eq.  (3.10) after replacing the l. h. s.  to $ \exp
(4i\n x_1) $.

There are the following keypoints, that make this solution
meaningful. The first is the finiteness  of
the intermediate region size $d$, because for an infinite region $II$
the solution (3.6) is obviously unacceptable.  In our case, however,
the size of the intermediate region is always finite by construction,
and the boson field $\vf (x)$ acquires the solitonic behavior in the
region $III$ due to the self-interaction $V(\vf)$. Here the following
circumstance manifests again: in (1+1)D the chiral coupling $\bp
e^{i\g_5 \vf} \p$ itself cannot cause the solitonic behavior of the
scalar field   by virtue of the effects of fermion-vacuum
polarization only, i.e. without additional self-interaction of bosons
[25].

Another important circumstance is the calculation of the Dirac sea
average of axial current $J_5$ by means of
the symmetry $ \n \leftrightarrow -\n $. The point is that for such
kind of averages another consistent definition is possible, namely,
via so called $\eta$-invariant, which is often used for the study of
fermion vacuum polarization effects [26]:  $$ \< J_5 \>_{sea}=\lim
\limits_{\eta \to 0} \[ \half \sum \limits_{\w_n <0} e^{- |\w_n| \eta
} \c^+_n i\g_1 \c_n- \half \sum \limits_{\w_n >0} e^{-\w_n \eta}
\c^+_n i\g_1 \c_n \] \ .  \eqno(3.18)$$ Since the spectrum of $\w_n$
is not symmetric, the expression (3.18) doesn't vanish, in
contrast to the average (3.17). The expression (3.18)  with
connection between $\w_n$ and $\n_n$ taken into account, transforms
into $$ \lim \limits_{\eta \to 0} \[ \sh \l \eta \( \sum
\limits_{\n_n >0} e^{-\n_n \eta} \c^+_n i\g_1 \c_n \) \] \ ,
\eqno(3.19)$$ and for $\eta \to 0$ the sum in (3.19) diverges as
$1/\eta$, so such $ \< J_5 \> $ appears to be proportional to $\l$
and doesn't vanish unless $\l \not = 0$.  The same makes the r.h.s.of
eq. (2.4b),   and so the linear functions will not be the solution of
this equation in the region $II$.

Moreover, this is true  for other fermionic averages too,
in particular for the v. e. v. of the fermion charge. In our
approach,  by virtue of discreteness of the fermionic spectrum, the
vacuum value of the charge is zero, in analogy to the case of
the axial current.  Namely, if  the $C$-odd expression for the charge
$$ Q= \half \int \! dx \  [ \p^+, \p]_- \ , \eqno(3.20)$$ is
used, then under the same conditions of sign-correspondence between
$\w_n$ and $\n_n$ for each $n$, one obtains for the average of $Q$
the following $$ \< Q \>=\< Q \>_{sea} =  \half \sum \limits_{\w_n
<0} -\half \sum \limits_{\w_n >0} = \half \sum \limits_{\n_n <0} -
\half \sum \limits_{\n_n >0} =0.  \eqno(3.21)$$ However, if $ \< Q
\>_{sea} $ is defined via $\eta$-invariant, then one obtains $$ \< Q
\>_{sea} = \lim \limits_{\eta\to 0} \[ \half \sum \limits_{\w_n <0}
e^{-|\w_n| \eta} - \half \sum \limits_{\w_n >0} e^{-\w_n \eta} \]
\not= 0 \ \eqno(3.22)$$ for $\l \not= 0$ by virtue of absence of
the symmetry $ \w \leftrightarrow -\w $.  Nethertheless, there are
serious reasons  to consider the relations (3.17) and (3.21) as the
most adequate way of calculation the sea average in our problem.
First, $\eta$-invariant is actually the measure of the Hamiltonian
asymmetry, rather than the fermion charge of the ground state.
Second,  the other regularization
schemes besides of thermal regularization are possible for divergent
sums such as $\< J_5 \>$ and $\< Q \>$. Also,
there is no warranty that the regularized expressions (3.18) and
(3.22) would produce the correct results for $\eta \to 0$.  It is
easy to give an example, when the dependence on a parameter in the
sum or integral is not continuous, in particular, $F(k)= \int
\nolimits _0^{\infty} \! dx \ \sin kx /x =\pi/2 $ for any $k > 0$,
but $F(0) =0$, i.e. in this case the limit of the integral for $k \to
0$ and it's exact value at $k=0$ are different.

Let us consider further an adiabatic process of changing the
gradient of the boson field $\l$ in the region $II$, starting from $\l =
0$.  In the initial moment $\w_n=\n_n$, so any $C$-odd fermionic sea
average obviously disappears: $\< . \ . \ . \ \>_{sea} \equiv 0$ for
any reasonable regularization scheme. Now let us take into account
that in our case the fermionic spectrum is discrete and
depends on $\l$ continuously, so for $\l$ small enough  all the
$\w_n$'s keep their signs.  Then from general grounds $\< . \ .
\ .  \ \>_{sea}$ should vanish still, although the spectrum $\w_n$
is already non-symmetric and so the $\eta$-invariant is nonzero.
  Therefore it is reasonable  to assume that in this case
the thermal regularization can yield physically inconsistent results
for $C$-odd quantities such as the axial current and the fermion
charge of the vacuum.  It is worth-while to note the
essential difference of this situation from the case of infinite
space, when the spectrum is continuous. In the latter case, the
density of states changes for any small change of $\l$, and so all
the averages change too, what gives rise to the effect of induced
fermion numbers [26].

Finally, it is possible to check the self-consistency of solution (3.6)
numerically. In this case the problem is solved on a lattice, and
so the number of degrees of freedom, as well as  the number
of fermion levels, is finite, so one doesn't need any regularization
at all.  The results of these calculations show that the linear
function (3.6) and the vanishing averages are the only
self-consistent solution to this problem.

The interpretation of this result is as following. In our case the
boson field has no discontinuities at all (it doesn't vanish in
the region $I$, too), and is topologically equivalent to the odd
soliton which would take place in absence of fermions due to the
self-interaction $V(\vf)$ only.  Therefore, the topological number
does not depend on the existence and sizes of the spatial regions with
fermions ($I$ and $II$).  On the other hand, the baryon charge of
the hybrid bag is, by definition, the sum of the topological charge of
the boson soliton and the fermion charge of the bag interior. The
latter one is zero in our case, hence the baryon charge is
determined completely by the topological charge of the boson field,
and so does not depend on the sizes of the regions $I$ and $II$, in
correspondence with the general ideology of hybrid models.
Therefore, in our approach the ``hadron'' is a composite particle,
which consists of the boson soliton, to which the fermion bag is
coupled through the interaction in the intermediate region $II$.
Note also, that although quantum numbers of such a composite particle
are completely determined  by the soliton, it doesn't mean that
the filled fermion levels with positive energy should not exist at
all.  This could take place for small enough values of the parameter
$\l$ only.  If $\l$ increases,  the negative levels $\w_n= - |\n_n| +
\l $ will unavoidably move into the positive part of the spectrum.
The change of sign of each such level will decrease $\<Q\>_{sea}$ by
one unit of charge, but if we fill the emerging positive level with
the valence fermion, then the sum $Q_{val}+Q_{sea}$  remains
unchanged.  Analogously, the total axial current will be equal to
$J_{val}+J_{sea}$ and will not change too, what ensures the vanishing
r.h.s. of (2.4b) and so preserves the status of linear function (3.6)
as the self-consistent solution of the field equations.  In other
words, by definition the ground state of the bag is a state in which
all the levels with $\n_n < 0$ are filled (the inequality is strong,
because there are no levels with $\n=0$ in the considered case).
The existence or absence of valence fermions in such
construction of the ground state of the bag depends actually on the
relation between $\l$ and $|\n_n|_{min}$, and so appears to be a
dynamical quantity like the other parameters of the bag (the size and
mass), which are determined from the total energy minimization
procedure.

\section{The Total Energy of the Bag}

On the unification  $I+II$ the boson soliton takes a form of the
linear function (3.6). This function (after rescaling $\vf \to \vf
/g$) is sewn together with the soliton solution of eq.  (2.5b) in
the exterior region by means of the continuity conditions for the
field and its derivative.  To preserve the generality of
consideration, in the region $III$ we'll use the asymptotic expansion
of the soliton solution of eq.  (2.5b) for large $|x|$, namely  $$
\vf_{sol}(x)={\pi \over g} \(1 - Ae^{-mx} \) \ ,  \quad x>x_2 \ ,
\eqno(4.1)$$  with $m$ being the meson mass in the exterior region of
the bag, while  for $x<-x_2 \quad \vf_{sol}(x) $ is determined by
oddness.  The factor $\pi / g$ means that we deal actually with a
phase soliton with topological charge being multiple of $2\pi/g$,
since it is the period of the initial chiral interaction $\bp \exp
(i\g_5 g\vf) \p$.  The constant $A$ is determined from the continuity
condition for boson field at points $x=\pm x_2$. Note that in the
exterior region there is no chiral invariance due to the
phenomenology of strong interactions on the one hand, and  on the
other due to the specific features of (1+1)D scalar models, which
make the presence of the  meson mass the necessary condition for  the
required soliton profile to be formed.

The continuity conditions at $x=\pm x_2$ give
$$ 2\l x_2 = \pi \left(1-Ae^{-mx_2}\right)
\eqno(4.2a)$$ $$ 2\l = \pi m A e^{-mx_2}. \eqno(4.2b) $$
Then one finds  the relation between the parameter $\l$ and the
bag's size $x_2$:  $$2 \l = \pi \frac{m}{mx_2+1}, \eqno(4.3)$$
whence the total energy of boson soliton can be represented as
$$E_{\vf} = {\pi^2 \over g^2} {m \over mx_2+1}.  \eqno(4.4)$$ The
total energy of the bag consists  of $ E_{\vf}$ and of the fermionic
contribution  $E_{\p}$ $$ E_{bag}= E_{\vf} + E_{\p} .  \eqno(4.5)$$
It is obvious from (4.4), that the energy of the boson field is a
smooth function, which decreases for $x_2 \to \infty$, producing no
vacuum pressure  in spite of the fact, that the gradient of $\vf$ in
the domain $I+II$ yields the constant positive contribution to the
energy density $\1/2 \vf'^2=2\l^2$, which could be identified with
the vacuum pressure $B$ in the standard HCM. Actually, it is an
artifact of one spatial dimension in our problem:  when the bag's
size increases, the gradient of $\vf$ in $I+II$ will always decrease
in any number of space dimensions by virtue of (4.3), while the
volume of domain $I+II$ in 1(space)D increases only linearly and
so cannot compensate the decreasing of $\l$, what takes place
in 2- and 3-(space)D. So in (1+1)D the non-trivial dependence
of the total energy $E_{bag}$ of the bag  on the model parameters
could  originate only from the fermion contribution $E_{\p}$, which
is given by the sum of the Dirac's sea of filled
negative energy states and positive energy  valence fermions
$$ E_{\p}= E_{val} +E_{sea} \ . \eqno(4.6)$$  For the ground state of
the bag described above, the sum (4.6) can be
reduced to the single universal expression by taking into account,
that the charge conjugation symmetry dictates the following
definition of the Dirac's sea energy [25,27]  $$ E_{sea}= \1/2 \sum
\limits_{\w_n < 0} \w_n - \1/2 \sum \limits_{\w_n > 0} \w_n \ .
\eqno(4.7)$$ If the  transition
from $\w_n$ to $\n_n$ is sigh-preserving for all $n$, and so there
are no valence fermions in the ground state of the bag (in order to
provide the vanishing v.e.v. for the charge and axial current), one
finds from (4.7) $$ E_{\p}=E_{sea}= \1/2 \sum \limits_{\n_n > 0} (-
\n_n + \l) - \1/2 \sum \limits_{\n_n > 0} ( \n_n + \l) = - \sum
\limits_{\n_n > 0} \n_n \ . \eqno(4.8)$$ If the parameter $\l$
appears to be large enough, so that the initially negative level
$\w_n=-|\n_n|+ \l $ changes its sign and turns into the filled
valence state, it is convenient to calculate $ E_{\p}$ in two steps.
First, we consider the contribution from all states with $|\n_n|>\l$
to $E_{sea}$, which in analogy to (4.8) reads $$ E'_{sea}= - \sum
\limits_{\n_m > \l }  \n_m \ .  \eqno(4.9)$$ To this expression the
energy of the valence fermions $E_{val}=-|\n_n|+ \l $ and the
  contribution of the positive levels with $\w_n=\pm |\n_n|+ \l $
should be added, what yields $$ E_{\p}=-|\n_n|+ \l - \1/2 [(-|\n_n|+
\l)+(|\n_n|+ \l)] + E'_{sea} = - \sum \limits_{\n_n > 0} \n_n \ ,
\eqno(4.10)$$ i. e. the same eq.  (4.8) as we have got for the energy
of fermions without filled valence states.

It is convenient to introduce a set of new parameters, in terms of which
the total energy of the bag will be expressed in the most appropriate
form.  First of all, we introduce the dimensionless quantities $$
\a=2Mx_1 \ , \quad  \b=2Md, \quad \r=2Mx_2 \ ,  \eqno(4.11)$$ and
 consider the eq. (3.10) in their terms.  This equation has two
branches of roots.  The first one corresponds to real $k$ and in
terms of parameters $\a$ and $\b$ is determined from $$ \tan \(\a
\sqrt{1+x^2}\) = { x \over \sqrt{x^2+1} } \ { x \cos \b x + \sin \b x
\over 1- \cos \b x + x \sin \b x } \ , \eqno(4.12)$$ where the
 unknown quantity is the dimensionless $x$ defined through $k=Mx$,
 $\n=M \sqrt{1+x^2}$.  The real roots $x_n$ belong to the
half-axis $0 \leq x_n < \infty $, since the fermionic wave
functions are actually the standing waves in a finite spatial box
with degeneracy in the sign of $k$, while the corresponding
frequencies $\n_n$ lye in the interval $M \leq \n_n < \infty$.

The second branch corresponds to  imaginary $k=iMx,
\ \n=M \sqrt{1-x^2} \ , \ 0 \leq x \leq 1$,
and can be derived from (4.12) by means of the analytical continuation
$$ \tan \(\a \sqrt{1-x^2}\) = { x \over \sqrt{1-x^2} }
 \ { x \ch \b x + \sh \b x \over \ch \b x + x \sh \b x -1 } \ .
 \eqno(4.13)$$ For this branch $0 < \n_n \leq M$.

Therefore, $\n_n$ and  $E_{\p}$ appear to be the functions of two
independent dimensionless parameters $\a$ and $\b$, the sum of which
defines  the dimensionless size  of the confinement domain $\r$
 $$ \a + \b = \r \ .  \eqno(4.14)$$ Proceeding further, it is
convenient to extract the mass of a ``constituent quark'' $M$ from
the sea energy and fermion frequencies as a dimensional
factor:  $$ \n_n=M \ve_n , \quad \ve_n=\sqrt{1+x_n^2} \ ,
\eqno(4.15)$$ hence $ E_{\p}= -M \sum \nolimits_n \ve_n$.
Introducing further the dimensionless ratio of the two model mass 
parameters $$ \m = m /2M \ , \eqno(4.16)$$ and the dimensionless 
total energy $\E_{bag}=E_{bag}/M $, for the latter one finds $$ 
\E_{bag} = \E_{\p} (\a, \b) + {\pi^2 \over g^2} { 2 \m \over \m \r+1 
} \ .  \eqno(4.17)$$ Let us note, that in this case the independent 
model parameters are $\a, \ \b$, which define the sizes of the inner 
bag's regions, while $\r$ is determined from (4.14). In this point 
  the present version of 3-phase model differs from the one 
  considered in [21], where $\a, \ \b$ are defined unambiguously by 
  $\m$ and $\r$.  So the total bag's energy  depends ultimately  on 
  the following dimensionless parameters:  $\m, \ g$ and $\a, \ \b$, 
where the parameter $\m$ is fixed by the ratio of the masses $m$ and 
$M$, while the optimal values of $\a, \ \b$ for the ground state 
should be found from the condition of minimum of the total energy for 
given $\m$.

The dimensionless fermion sea
energy $\E_{\p}$, which obviously diverges in the upper limit, should
be regularized, so  a consistent renormalization procedure is
required. First, let us consider the asymptotics of roots of eq.
(4.12) in the UV-domain, when $x_n \gg 1$. It is convenient to
rewrite the eq.  (4.12) as $$ \sin \a \sqrt{1+x^2} = $$ $$ = \1/2
(\sqrt{1+x^2} +x ) \sin \( \a \sqrt{1+x^2} + \b x + \g \) + \1/2
(\sqrt{1+x^2} -x ) \sin \( \a \sqrt{1+x^2} - \b x - \g \) \ ,
\eqno(4.18)$$ where $\g = \arctan x \ . $ One can show that $$
x_n(\a, \b) = { \pi/2 + \pi n \over \r} + { (-1)^{n+1} \sin \[ (\pi/2
+ \pi n ) \a/\r \] +1 - \a/2 \over \pi /2 + \pi n } + O (1/n^2) \ .
\eqno(4.19)$$ In the expression (4.19) the first term leads to the
quadratic and linear divergences in $\sum \nolimits_n \ve_n $, and
the second one contains the logarithmic divergence, while the term
with the sine does not yield any divergence at all.
Therefore, the UV-regularization consists in the
compensation of the first term and the divergent part of the second
term in the asymptotic expression (4.19). As a first step we use that
obvious fact, that only the difference between two energies is
physically meaningful, rather than the energy itself.  It could seem,
that the most natural way is to choose the energy  of the free
fermions sea in the similar ``volume'' $\r$ as the reference point
for $\E_{\p}$.  However, it has been shown in [21], that this
subtraction fails in our case because this energy appears to be
infinitely larger than any configuration with $\b \not= 0$, even
after any possible counterterms are added.   Such an infinite energy
barrier between the regularized $\E_{\p} (\a, \b)$ and the free
fermion sea corresponds perfectly well to the intuitive feeling that
the free fermions can hardly be a reasonable first approximation to
the confinement problem.

As a result, in this case there is no  unique
prescription for the  choice of subtraction point in the
renormalization of $\E_{\p} (\a, \b)$, what is actually the common
feature for the majority of bag models [10,12,28].  In the
``classical'' renormalization procedure, the uncertainty in the
 choice of subtraction point is cancelled by  fixing the physical
values for corresponding number of parameters. For obvious reasons,
we won't do that in our ``toy'' (1+1)D model, but instead will
consider the most  straightforward approach to the compensation of
divergences in the sum (4.10), which preserves the continuous
dependence of the result of subtraction on the model parameters.
This approach is based on the subtraction from $\sum \nolimits_n
\ve_n$ the other sum with the same summation index $n$, the common
term of which coincides exactly with the divergent part of
asymptotics (4.19), resulting in the finite quantity $$ \tilde
{\E}_{\p}= -\sum \limits_n \[ \ve_n - \( {\pi/2+\pi n \over \r}+ {1
+\b/2 \over \pi/2 + \pi n} \) \] \ .  \eqno(4.20)$$ This method
requires no counterterms because all the  divergencies are already
cancelled by the subtracted sum. Of course, to some extent the
physical meaning of such procedure is lost.  However, it should be
emphasized, that it is only the (1+1)D case, when the theory with
coupling $\lagr_I=G \bp (\s +i\g_5 \pi )\p$ is (super)renormalizable
and any counterterm has explicit physical meaning.
For higher space dimensions this is already not true, and so the
procedure of compensation of divergences in the energy based on
(4.20) should not be considered as having no motivation.
For more detailed discussion on the extraction of finite part from
the divergent Dirac's sea energy see refs. [28-31].

Proceeding further, let us turn to  the study of the total bag's
energy $$ \E_{bag}=\tilde {\E}_{\p}(\a,\b) +  {\pi^2 \over g^2} \ {
2\m \over \m\r+1 } \eqno(4.21)$$ as a function of parameters $\a,
 \ \b$.  The analysis
of the contribution of convergent logarithmic part from the sine-term
in the asymptotic expression (4.19) to $\tilde {\E}_{\p}$ yields the
first feature of $\E_{bag}$.  Let us transform this contribution
to the form $$ \(\tilde {\E}_{\p}\)_{log} (\a,\b) = { 1 \over \pi}
\sum \limits_{n \gg 1} (-1)^{n} { \sin \[ (\pi \a / \r) ( n + 1/2 )
\]  \over  n + 1/2  } \ \eqno(4.22)$$ and then use the wellknown
relation $$ \sum \limits_{n=0}^{\infty}  (-1)^{n} { \sin \[ z ( n +
1/2 ) \] \over  n + 1/2  } = \ln \tan (\pi/4+z/4) \ .  \eqno(4.23)$$
It is easy to see, that the sums (4.22) and (4.23) have the similar
common term, and   the sum (4.23) diverges as $- \ln
(\pi - z)$ when $z \to \pi$. Hence, for $ \pi \a / \r \to \pi   $ ,
what means  either $\b \to 0$, or  $\a \to \infty$ for finite
$\b$, the sum (4.22) will show the similar behavior, namely  $$ \(
\tilde {\E} _{\p}\)_{log} (\a,\b) \to - {1 \over \pi} \ln (\b/\a) \ ,
\quad \b/\a \to 0 .  \eqno(4.24)$$ Therefore, both the regularized
fermion energy (4.20) and the total bag's energy
possess the logarithmic singularity for $\b \to 0 $, and at the
same time the logarithmic increase for  $\a \to \infty$ and finite
$\b$.  Note that although the possibility of smooth transition to
a 2-phase configuration, when $d \to 0$, exists formally for the
initial Lagrangean (2.2), actually it disappears due to the
singularity for $\b \to 0$. In other words, in such 3-phase model the
radial extent of the intermediate phase can be sufficiently small,
but not zero, what corresponds to the general physical treatment of
the structure of many-phase systems.

$\E_{bag}$ will also grow for $\b \to \infty$ and finite $\a
$.   Because the increase of $\b$ yields the increase of
$\r$, in this case $ \pi \a / \r \to 0$, hence the logarithmic term
(4.22) becomes negligibly small, and the main contribution is
provided by the next terms in expansion in $1/n$.  However, it is
technically more convenient to use the fact that for $\r \to \infty$
the fermionic spectrum becomes quasicontinuous with the exception of
a small vicinity of Fermi surface with zero energy, what allows to
transform the sums over $x_n$ into integrals over $dx$.  In
particular, the analysis of  distribution of the  roots of eq. (4.12)
shows, that for this limit $\sum_n \ve_n$ is approximated by the
following (divergent) integral $$\sum_n \ve_n \to {1 \over \pi} \
\int \! dx \ \sqrt{1+x^2} \ \times $$ $$ \times \[ \b + {1 \over
1+x^2} + \a {x^2 \over x^2+ \sin^2 \(\a\sqrt{1+x^2}\)} - {
\sin\(\a\sqrt{1+x^2}\) \cos\(\a\sqrt{1+x^2}\) \over  \sqrt{1+x^2} \(
x^2+ \sin^2 \(\a\sqrt{1+x^2}\) \) } \] \  .  \eqno(4.25)$$ For the
subtracted sum in (4.20) one can easily find $$ \sum \limits_n
\({\pi/2+\pi n \over \r}+ {1 +\b/2 \over \pi/2 + \pi n} \)  \to  {\r
\over \pi} \ \int \! dx \ \( x+ {1+\b/2 \over \r x} \)  \ .
\eqno(4.26)$$ The integrals (4.25) and (4.26) have the same divergent
part $$ {1 \over \pi} \ \int \! dx \ (\r x + 1/x + \b /2 x) \ , $$ so
their difference yields a converging integral, in agreement with the
subtraction procedure. The leading term of the intergrand in  this
difference, taken with the (correct) inverse sign,  is $\b/8\pi x^3$.
This finally leads to the emergence of the positive, proportional to
$\b$, contribution to $\tilde {\E}_{\p}$, and correspondingly to
$\E_{bag}$.

Finally,  for $\a \to 0$ and finite $\b$ there is
no singularity  in $\tilde {\E}_{\p}$ and $\E_{bag}$ at all, what is a
direct consequence of the analysis performed in ref.[21], where
it has been shown that the renormalized $\tilde {\E}_{\p}$'s for
$\a=0$ and finite $\a \not =0$ differ always by a finite value.
Therefore it is the soliton energy  $\E_{\vf}$ and so the
current values of parameters $\m$ and $g$, which play  the main role
in this domain. More exactly,  for $\m$ being not too small, the
dependence of $\E_{\vf}$ on $g$ will be such that for sufficiently
small $g$ the sharp enough increase of $\E_{bag}$ takes place for $\a
\to 0$ and finite $\b$, providing the existence of a well-defined
minimum.  On the contrary, the sufficient increase of $g$ (or $\m \to
0$) could make $\E_{\vf}$ to be negligibly small (and almost
constant) for any  bag's size, and then the minimum in the total
energy disappears.

The numerical calculation confirms completely such qualitative
predictions for the behavior of $\tilde {\E}_{\p}$  and $\E_{bag}$.
However, due to the absence of singularity in $\E_{bag}$ for $\a = 0$
the question about the existence of the minimum of the total bag's
energy considered as a function of $\a,\b$, and therefore the
existence of a stable ground state of the bag itself for given $\m$
and $g$ can be answered only numerically.  Such
calculation has been performed for $\m=0.25$, what corresponds
approximately to the ratio $m_{\pi}/2m_Q$, where the constituent
quark mass is assumed to be equal to 300 MeV, and for $g=1$.  The
qualitative behavior of $\E_{bag}$ as a function of $\a , \b$ is
shown on the Fig.1, from which one can easily see, that the total
energy reveals the unique minimum for non-zero values of $\a , \b
$.  On the Fig. 2 the  equal energy  curves are shown, what allows
to observe this minimum of energy more explicitly. In the case under
consideration the values of $\a$ and $\b$ in the minimum differ by
two orders, i.  e.  the size of the intermediate region appears to be
essentially smaller than that of the inner one. Therefore to some
extent it can be treated as a smeared boundary
between the phase of asymptotic freedom and the purely colorless
phase. On the Fig. 3 the scheme of fermionic levels in the vicinity
of zero energy is presented for this configuration, which shows that
in this case  the ground state contains one  filled valence level
with positive energy.

\section{Conclusion}

The aim of the present study is the construction of a consistent model
of a hybrid chiral bag, in which the exact equivalence between fermionic and
bosonic (meson) phase is not assumed. Our results show that
such a model can be actually formulated in a quite consistent
manner, and to certain extent could be more effective way of
description of low-energy hadron physics compared  to the
traditional HCM.

First,  let us note, that the initial formulation of the model is a
local field theory, and in spite of the variety of classical
solutions one needs to deal with, the covariance is broken only
spontaneously, and so can be restored by means of the methods of
refs. [23] using the covariant group center-of-mass variables for a
localized quantum-field system.  Besides this, the positive points of
this approach are: the more correct derivation of chiral boundary
conditions, by which any term in  the initial Lagrangian has exact
physical meaning; the presence of an intermediate phase describing
quasifree massive ``constituent'' quarks; physically acceptable
behavior of the total energy of the bag as a function of it's
geometry.  Moreover, in this model the condition of fermion
confinement, incorporated into it from the very beginning, shows up
 more explicitly.  It manifests, in particular, in the fact
that there is no need in the term with vacuum pressure $B$,
which in the standard approach is inserted into the model by means of
some extra assumptions, since in our case the Dirac's sea
polarization itself produces the infinite increase of energy at
large distances.

The important question of the choice of
method of calculation of the Dirac's sea averages for fermion bags
should be also emphasized.  The method we used is based on the
discreteness of the fermion energy spectrum, what by
means of quite obvious considerations leads to very simple
solution of the self-consistent equations of the bag in the
intermediate region.  Note, however, that in spite of arguments in
favor of such method of calculations, we cannot completely reject
alternative methods like thermal regularization. The question of
which one is more adequate to the physics of the problem should be
answered only by means of detailed study of realistic models.

Compared to ref.[21], the specific feature of the considered
model is  that the condition of nonvanishing density of the boson
condensate in the interior region affects crucially the amount of
possible bag configurations, which provide the local minima of the
energy. In ref. [21]  the boson field vanishes in the inner region,
what  leads to  an infinite set of such configurations with the same
topology and infinitely increasing size and energy with the main
difference between them being the value of gradient $\l$ of the boson
field in the intermediate region.  In the present case, there exists
not more than one such a configuration, and the excited bag states of
greater energy can be obtained only by adding a number of valence
fermions.  So within the framework of 3-phase modification of
the hybrid model there are possible  quite different versions of
description of composite particles like hadrons and their
excitations.

\section{Acknowledgments}
This work was supported in part by RFBR
under Grants No. 96-02-18092, 96-15-96674, and Sankt-Peterburg
Concurrency Center of Fundamental Sciences, Grant No. 97-0-6.2-28.
One of authors (I. Ch.) thanks the Abdus Salam ICTP in Trieste for
the warm hospitality during Oct-Dec of 1999 where a part of this work
was done.

\section{Figures}
Fig. 1 \ The total energy of the bag $\E_{bag}(\a, \b)$ asM
a function of dimensionless bag  parameters $\a, \b$ for $\mu = 0.25$M
and $g=1$.

\noindent
Fig. 2 \ The curves of equal energy for $\E_{bag}(\a, \b)$M
for $\mu =0.25$.  The minimum is clearly seen  for $\a \simeq 40$M
and  $\b \simeq 0.34$. The values of energy in the minimum areM
$\E_{bag}=2.41, \ \E_{\p}=1.96$ and $\E_{\vf}=0.45$.

\noindent
Fig. 3 \ The scheme of fermionic levels for the groundM
state bag configuration.

\eject

\epsfbox{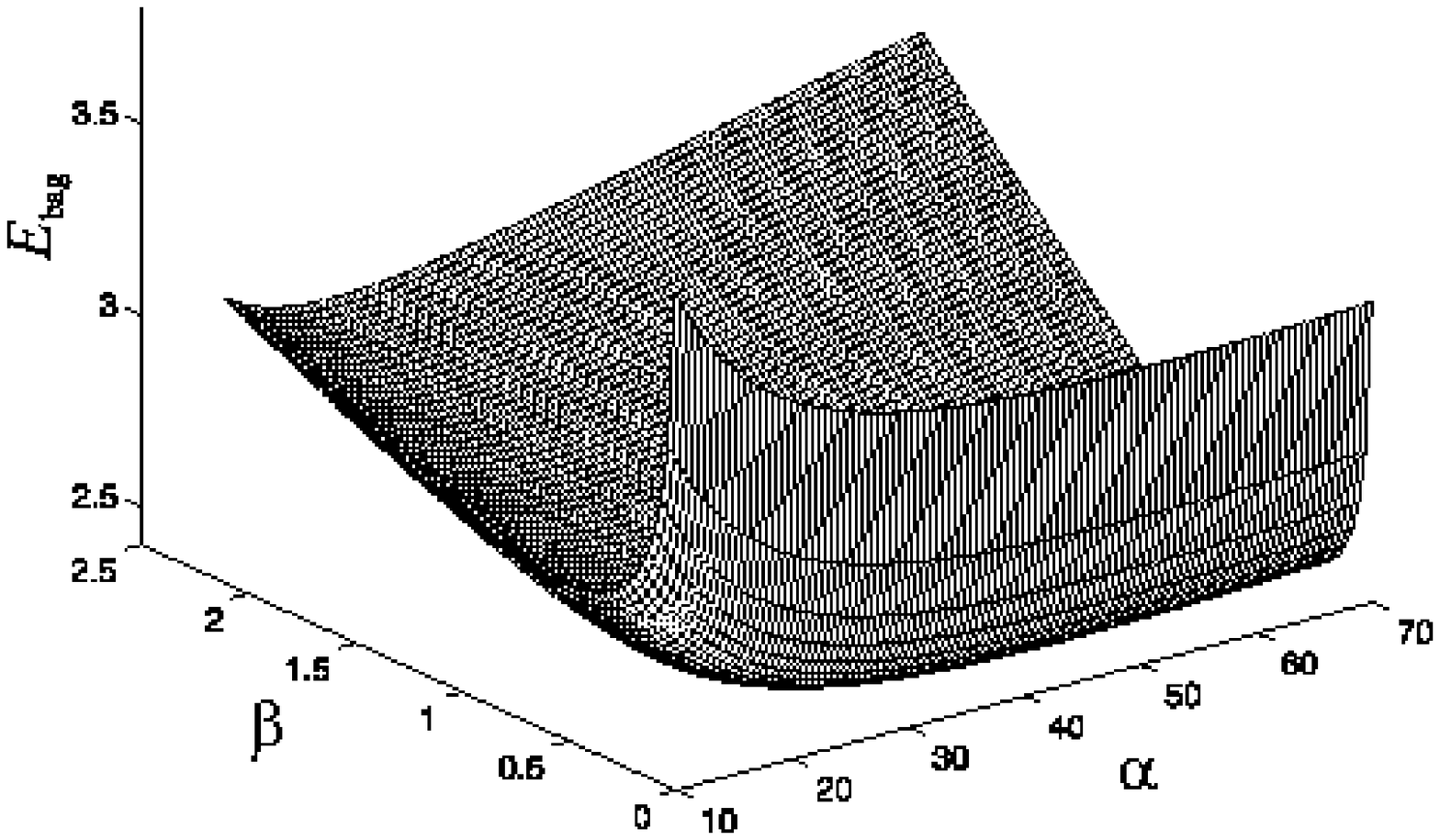}

\epsfbox{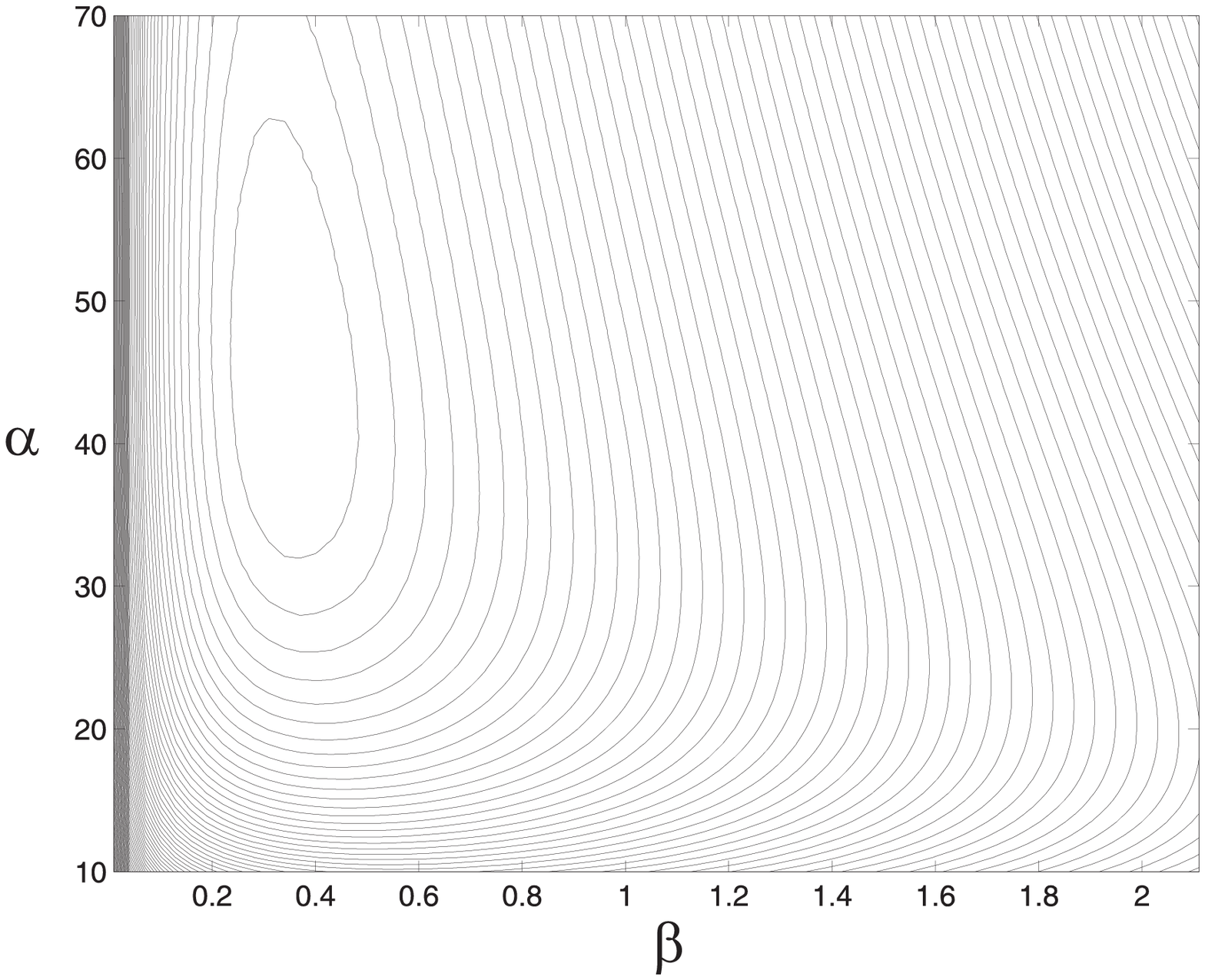}

\epsfbox{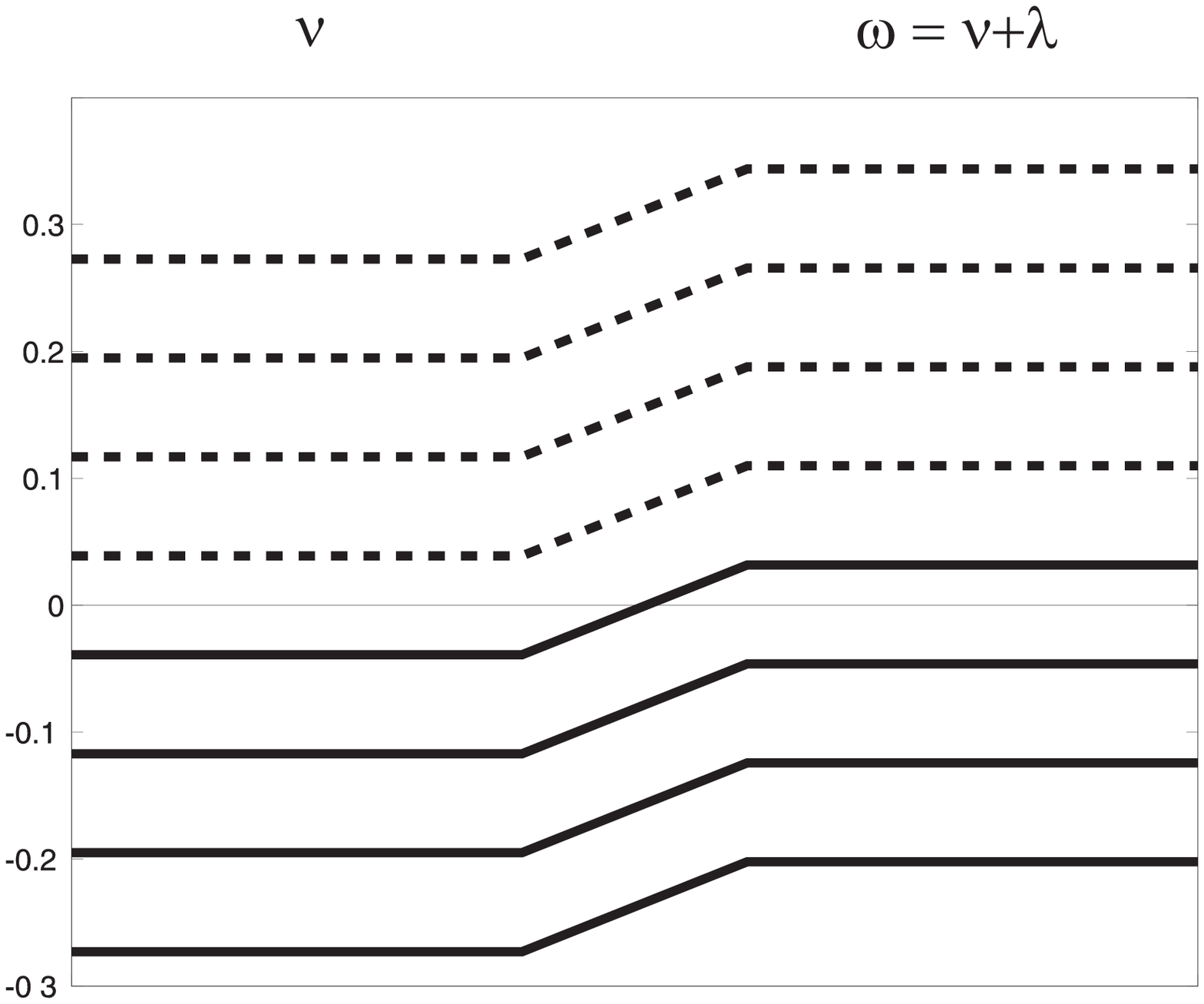}

\end{document}